%% file: arxiv.tex
\documentclass[12pt]{article}
\usepackage{epsfig,cite,amsmath,amssymb,color}

\graphicspath{{figs/}}

\evensidemargin \oddsidemargin
\newcommand{\mr}{\mathrm}
\newcommand{\bb}{$b\bar{b}$}

\newcommand{\ifb}{$\mbox{fb}^{-1}$}
\newcommand{\sixoo}{60 \ifb}
\newcommand{\sixooeff}{60 \ifb\,eff$\times2$}
\newcommand{\sixooo}{600 \ifb}
\newcommand{\sixoooeff}{600 \ifb\,eff$\times2$}
\newcommand{\Mhmax}{$M_h^{\mr max}$}
\newcommand{\BC}{\begin{center}}
\newcommand{\EC}{\end{center}}
\newcommand{\cp}{{\cal CP}}

\makeatletter
\renewcommand{\section}{\@startsection {section}{1}{\z@}%
                           {-3.5ex \@plus -1ex \@minus -.2ex}%
                           {2.3ex \@plus.2ex}%
                           {\mathversion{bold}\normalfont\Large\bfseries}}
\renewcommand{\subsection}{\@startsection{subsection}{2}{\z@}%
                           {-3.25ex\@plus -1ex \@minus -.2ex}%
                           {1.5ex \@plus .2ex}%
                           {\mathversion{bold}\normalfont\large\bfseries}}
\renewcommand{\subsubsection}{\@startsection{subsubsection}{3}{\z@}%
                           {-3.25ex\@plus -1ex \@minus -.2ex}%
                           {1.5ex \@plus .2ex}%
                           {\mathversion{bold}\normalfont\normalsize\bfseries}}
\makeatother

\begin{document}

\input{titlepage}



\section{Introduction}\label{Intro}
The central exclusive production of new particles has received a great deal of 
attention in recent years (see \cite{FP420TDR} and references therein). 
The process is defined as $pp\rightarrow p\oplus\phi\oplus p$
and all of the energy lost by the protons during the interaction
(a few per cent) goes into the production of the central system, $\phi$. The 
final state therefore consists of a centrally produced system (e.g. dijet, 
heavy particle or Higgs boson) coming from a hard subprocess, two very forward 
protons and no other activity. The '$\oplus$' sign denotes the regions devoid 
of activity, often called rapidity gaps. Studies of the Higgs boson produced in
CEP used to form a core of the physics motivation for upgrade projects to 
install forward proton detectors at 220~m and 420~m from the ATLAS 
\cite{ATLASTDR} and CMS \cite{CMSTDR} detectors, see \cite{FP420TDR}. At the 
moment, only 220~m stations are considered to be installed in ATLAS 
\cite{AFPLOI}. Proving that the detected central system is the Higgs boson 
coming from the SM, MSSM or other BSM theories will require measuring precisely 
its spin, \cp\ properties, mass, width and couplings.  

In \cite{HKRSTW} we have presented detailed results on signal and background 
predictions of CEP production (based on calculations in \cite{KMR} and the 
FeynHiggs code \cite{feynhiggs,mhiggsAEC}) of the light ($h$) and heavy ($H$) 
Higgs 
bosons which have then been updated in \cite{HKRTW}. Changes between these two 
publications are described in \cite{HKRTW} and summarized in \cite{ProcDIS11}. 
Four luminosity scenarios are 
considered: ``\sixoo'' and ``\sixooo'' refer to running at low and high 
instantaneous luminosity, respectively, using conservative assumptions for the 
signal rates and the experimental efficiencies (taken from \cite{CMS-Totem}); 
possible improvements on the side of theory and experiment could allow for 
scenarios where the event rates are enhanced by a factor 2, denoted by 
``\sixooeff'' and ``\sixoooeff''.

\section{Results and LHC exclusion regions}
\begin{figure}[htb]
\includegraphics[width=0.5\textwidth,height=5.2cm]{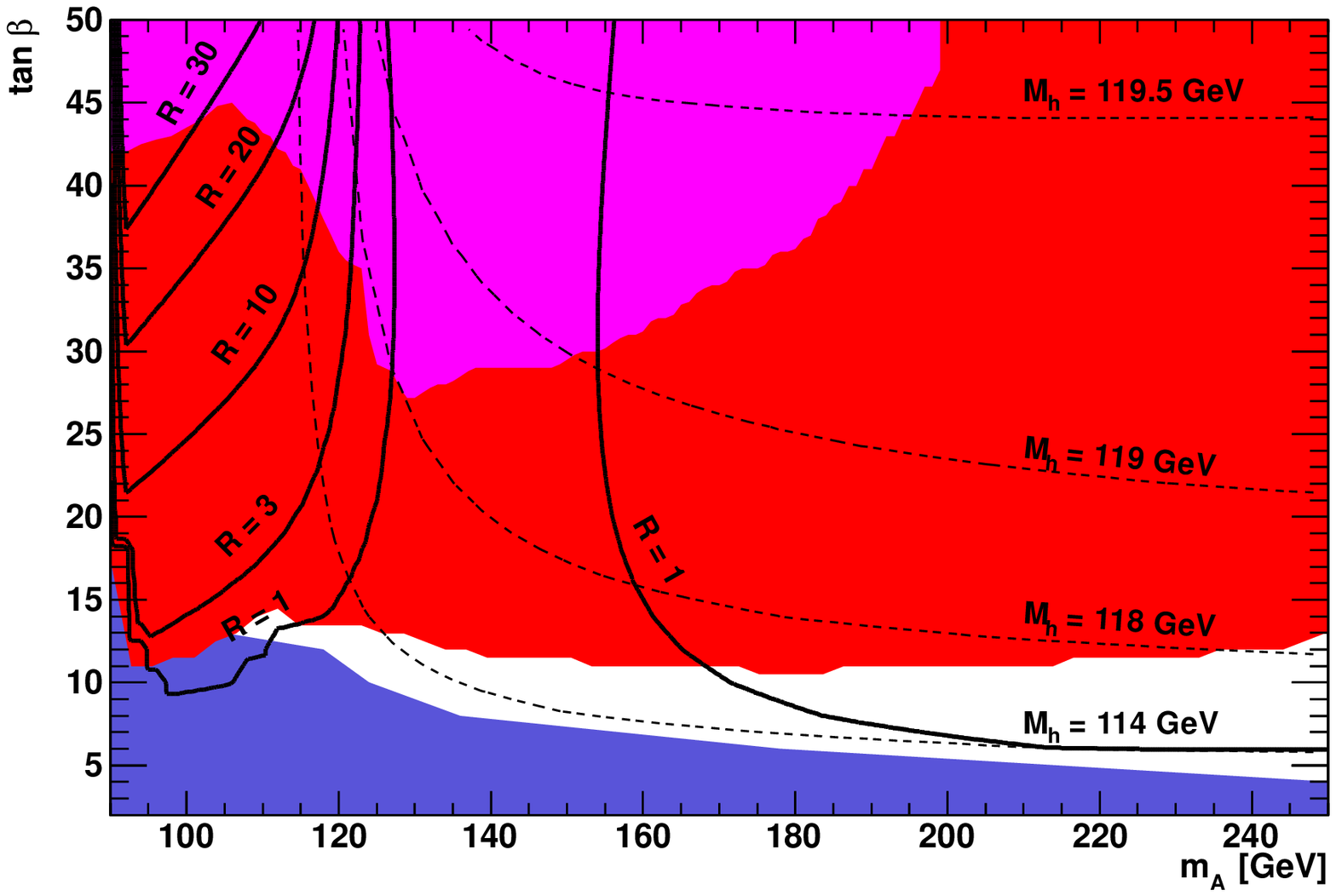}
\includegraphics[width=0.5\textwidth,height=5.2cm]{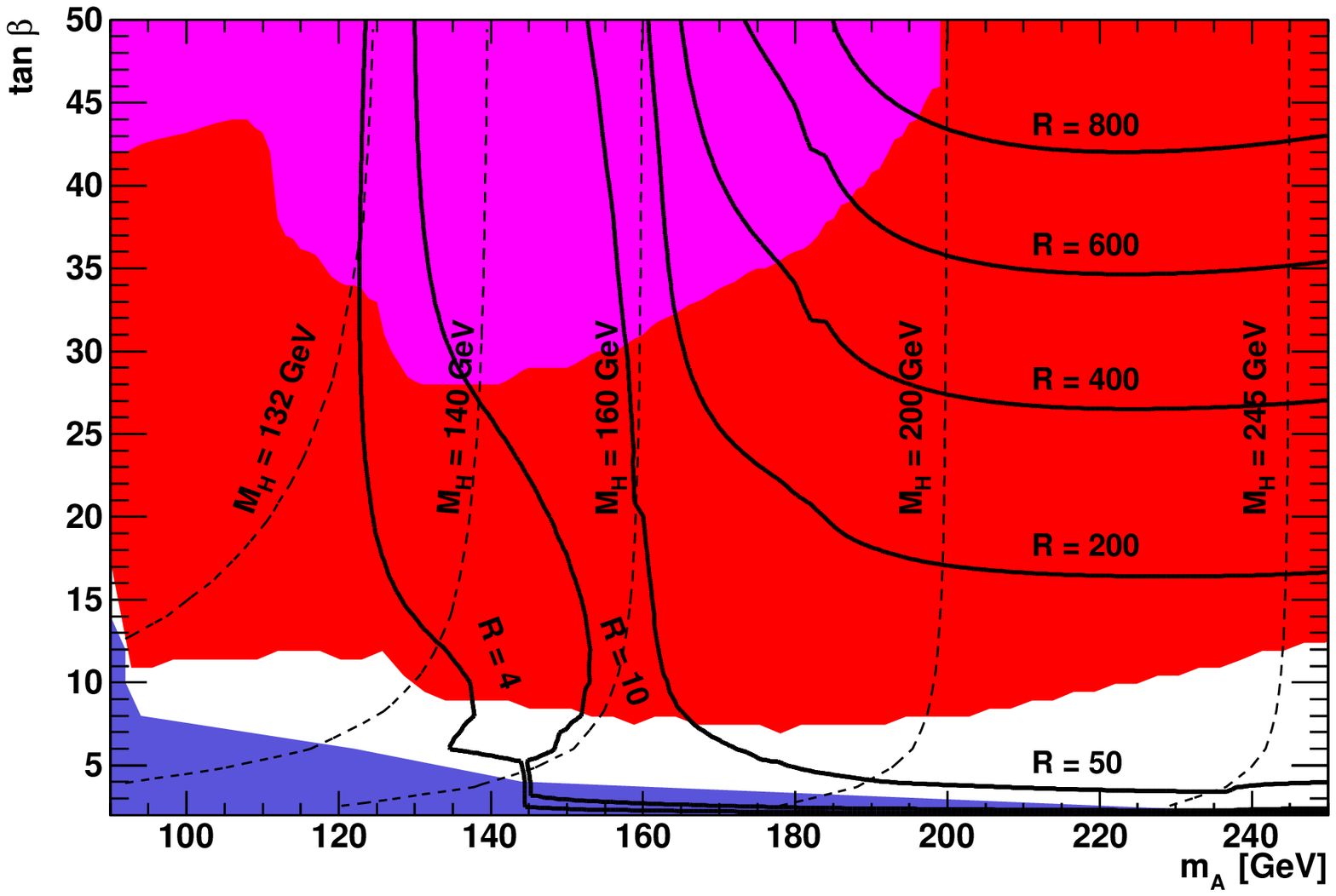}
\caption{Contours for the ratio of signal events in the MSSM to those in the SM
and for the mass values $M_h$ ($M_H$) for $h(H)\rightarrow$ \bb\ channel in CEP
in the $M_A-\tan\beta$ 
plane of the MSSM are shown on left (right) within the no-mixing (\Mhmax) 
benchmark scenario. The lighter (dark) shaded region corresponds to the 
parameter region excluded by the LEP (Tevatron) Higgs boson searches, the 
smaller region on top corresponds to the Tevatron Higgs boson search.}
\label{ratios}
\end{figure}

Standard benchmark scenarios designed to highlight specific characteristics of 
the MSSM Higgs sector, so called \Mhmax\ and no-mixing scenarios, do not 
necessarily comply with other than MSSM Higgs sector constraints. Scenarios 
which fulfill constraints also from electroweak precision data, B physics data 
and abundance of Cold Dark Matter (CDM) are the so called CDM benchmark 
scenarios \cite{CDM}. As observed and discussed in \cite{HKRTW}, the $5\sigma$
discovery and $3\sigma$ contours show in general similar qualitative features
as the results in the \Mhmax\ and no-mixing scenario. Since publications 
\cite{HKRTW} and \cite{ProcDIS11}, the results have been updated by adding the
exclusion regions coming from LHC searches for MSSM signal (see 
Fig.~\ref{ratios}). These exclusion regions are obtained using the code 
HiggsBounds version 3.6.0 \cite{higgsbounds}. Compared to previous results with 
Tevatron exclusion regions (\cite{HKRTW,ProcDIS11}), the allowed region for 
MSSM has now significantly shrunk.

\section{Hypothesis of a Higgs boson at 125~GeV}\label{hypothesis}
Presently a big effort is put into Higgs boson searches at the LHC, both at SM 
and beyond SM. While the MSSM exclusion regions are already accounted for in our
results (see e.g. Fig.~\ref{ratios}), the results of the SM Higgs boson search
on the data samples collected in 2011 (up to 4.9 \ifb) are 
as follows: both ATLAS (\cite{Atlas-Higgs}) and CMS (\cite{CMS-Higgs}) exclude
similar mass regions and both observe an excess of the signal over background 
in the same mass region. After combining all decay channels, ATLAS (CMS) declare
the excess at $M_h$=126~GeV (125~GeV) with a local 
significance of 2.5$\sigma$ (2.8$\sigma$). The global probability
for such an excess found in the search range 110 $< M_h <$ 600~GeV, in
the absence of a signal, is 2.2$\sigma$ (2.1$\sigma$). 
A natural question then is how 
the observation of Higgs candidates in this mass range would affect our results.
Let us work with a hypothesis that Higgs candidates are found at the mass of 
125 $\pm$ 3~GeV (1.5~GeV corresponds to the 
experimental uncertainty thanks to the fact that most of the signal comes from 
the $\gamma\gamma$ decay channel, being the most precise in the mass 
measurement; the theory uncertainty is estimated in \cite{theounc}). This SM 
Higgs mass range 122 $< M_h <$ 128~GeV is compatible with the allowed 
mass range 122.5 $< M_h <$ 127.5~GeV when combining exclusion limits 
found at 95\% C.L. by both experiments. The effect of this hypothesis is shown 
in Fig.~\ref{hypo125GeV} (now with the y-axis in the logarithmic scale) from 
which two main facts may be drawn: 
i) CEP MSSM signal still survives the as yet provided exclusion limits. 
In the allowed region, a significance of 
3$\sigma$ may be achieved for the highest luminosity scenario, ii) MSSM 
is in agreement with the tentative hints at 125~GeV, although the 
allowed region may shrink further with time.  

\begin{figure}[htb]
\includegraphics[width=0.5\textwidth,height=5.2cm]{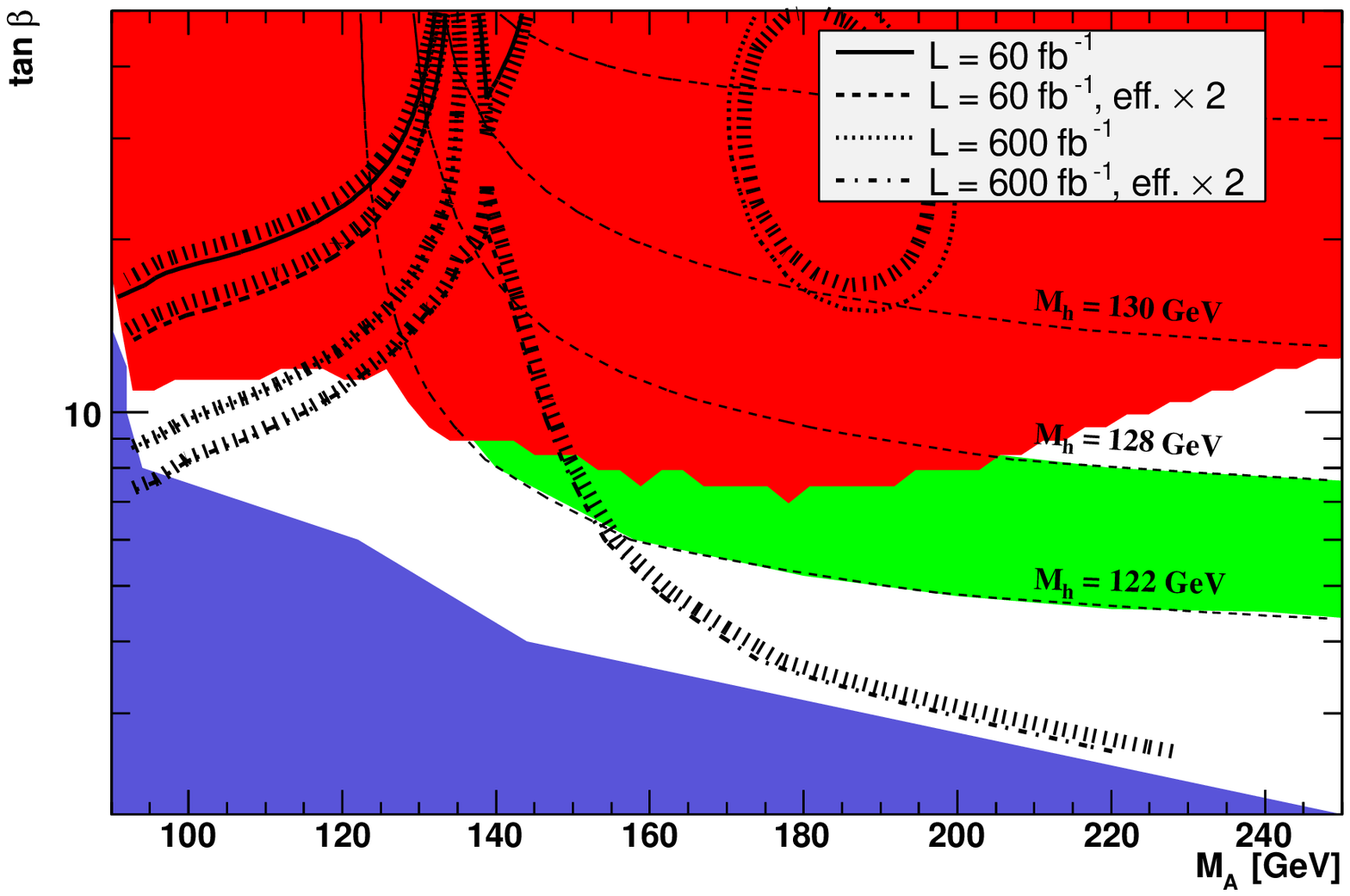}
\includegraphics[width=0.5\textwidth,height=5.2cm]{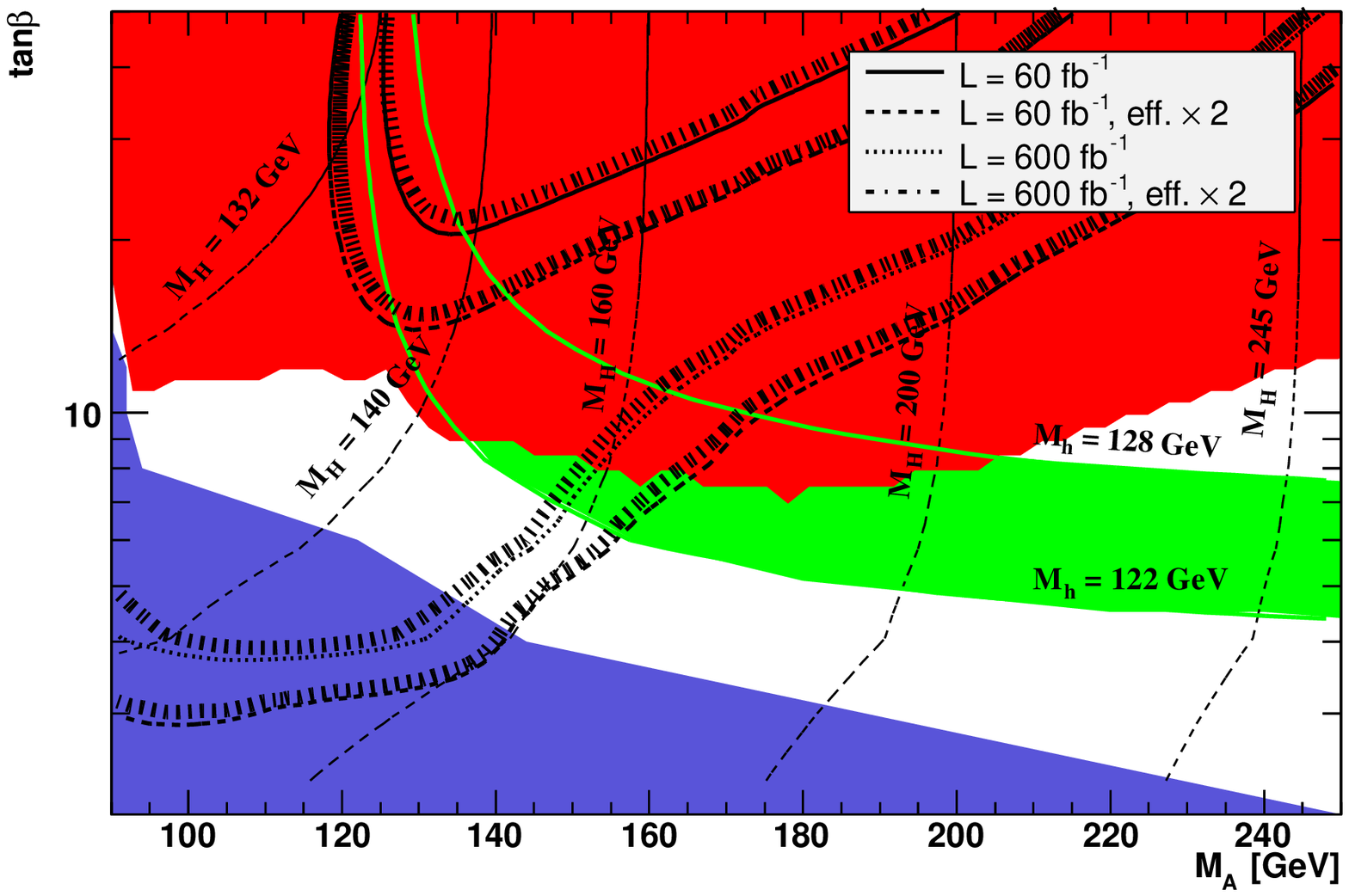}
\caption{3$\sigma$ evidence and mass $M_h$ ($M_H$) contours for 
$h(H)\rightarrow$ \bb\ channel in CEP production in the $M_A-\tan\beta$ plane 
of the MSSM are shown on left (right) within the \Mhmax\ benchmark scenario. 
The results are shown for four assumed effective luminosities (see the text). 
The lighter (dark) shaded region corresponds to the parameter region excluded 
by the LEP (Tevatron) Higgs boson searches. The region 122 $< M_h <$ 
128~GeV refers to the hypothesis of Higgs bosons found at 
125~GeV with assumed theory and experimental uncertainties.}
\label{hypo125GeV}
\end{figure}

\section{Coupling structure and spin-parity determination}
Standard methods to determine the spin and the \cp\ properties of Higgs bosons 
at the LHC rely to a large extent on the coupling of a relatively heavy Higgs 
boson to two gauge bosons. In particular, the channel $H\rightarrow 
ZZ\rightarrow\!4l$ - if it is open - offers good prospects in this respect
\cite{HZZ}. In a 
study \cite{Ruwiedel} of the Higgs production in the weak vector boson 
fusion it was found that for $M_H=160$~GeV the $W^+W^-$ decay mode allows the 
discrimination between 
two extreme scenarios of 
a pure \cp-even (as in the SM)
and a pure \cp-odd tensor structure at a level of 4.5--5.3$\sigma$ using about
10~\ifb\ of data (assuming the production rate is that of the SM, which is in 
conflict with the latest search limits from the Tevatron \cite{Tevlimits}).
A discriminating power of 2$\sigma$ was declared in the $\tau^+\tau^-$ decay 
mode at $M_H=120$~GeV and luminosity of 30~\ifb.

The situation is different in MSSM:
for $M_H \approx M_A \gtrsim 2 M_W$ the lightest MSSM
Higgs boson couples to gauge bosons with about SM strength, but its mass
is bounded to a region $M_h \lesssim 135$~GeV \cite{mhiggsAEC},
where the decay to
$WW^{(*)}$ or $ZZ^{(*)}$ is difficult to exploit.
On the other hand, the heavy MSSM Higgs bosons decouple from the 
gauge bosons. Consequently, since the usually quoted results for the 
$H \rightarrow ZZ/WW \rightarrow 4l$ channels assume a relatively 
heavy ($M_H \gtrsim 135$~GeV) SM-like Higgs, these results are not
applicable to the case of the MSSM. The above mentioned analysis of the weak
boson fusion with $H\rightarrow\tau^+\tau^-$ is applicable to the light 
\cp-even Higgs boson in MSSM but due to insignificant enhancements compared 
to the SM case no improvement can be expected. 

An alternative method which does not rely on the decay into a pair of gauge
bosons or on the production in weak boson fusion would therefore be of great
interest. Thanks to the $J_z=0$, C-even, P-even selection rule, the CEP Higss 
boson production in MSSM can yield a direct information about spin and \cp\
properties of the detected Higgs boson candidate. It is also expected, in 
particular in a situation where a new particle state has also been detected in 
one or more of the conventional Higgs search channels, that already a small 
yield of CEP events will be sufficient for extracting relevant information on 
the spin and $\cp$-properties of the new state \cite{HKRTW}.


\subsection*{Acknowledgments}
The work of M.T. was supported by the projects LA08032 and LA08015 of the 
Ministry of Education of the Czech republic.
The work of S.H. was supported in part by CICYT (grant FPA 2010--22163-C02-01) 
and by the Spanish MICINN's Consolider-Ingenio 2010 Program under grant 
MultiDark CSD2009-00064.

\end{document}

%% file: titlepage.tex
\thispagestyle{empty}
\setcounter{page}{0}
\def\thefootnote{\fnsymbol{footnote}}

\begin{flushright}
\mbox{}
DCPT/12/72\\
IPPP/12/36 \\
\end{flushright}

\vspace{1cm}

\begin{center}

{\large\sc {\bf Exclusive production of the MSSM Higgs bosons at the LHC}}
\footnote{talk given by M.T.\ at the {\em DIS 2012}, 
March 2012, Bonn, Germany}

\vspace{1cm}

{\sc 
S.~Heinemeyer$^{1}$%
\footnote{
email: Sven.Heinemeyer@cern.ch}%
, V.A.~Khoze$^{2}$%
\footnote{
email: V.A.Khoze@durham.ac.uk}%
,\\[.5em] M.~Tasevsky$^{3}$%
\footnote{
email: Marek.Tasevsky@cern.ch}%
~and G.~Weiglein$^{4}$%
\footnote{email: Georg.Weiglein@desy.de}
}

\vspace*{1cm}

{\it
$^1$Instituto de F\'isica de Cantabria (CSIC-UC), 
Santander,  Spain\\

\vspace{0.3cm}
$^2$IPPP, Department of Physics, Durham University, 
Durham DH1 3LE, U.K.\\



\vspace{0.3cm}
$^3$Institute of Physics, ASCR, 
Na Slovance2, 18221 Prague, Czech Republic

\vspace{0.3cm}

$^4$DESY, Notkestra\ss e 85, D--22607 Hamburg, Germany
}
\end{center}

\vspace*{0.2cm}

\BC {\bf Abstract} \EC
We review the prospects for Central Exclusive Production (CEP) of BSM Higgs 
bosons at the LHC using forward proton detectors proposed to be installed at
220~m and 420~m from the ATLAS and/ or CMS. Results are presented for MSSM in 
standard benchmark scenarios and in scenarios compatible with the Cold Dark 
Matter relic abundance and other precision measurements. Following results of 
the LHC Higgs boson searches, we investigate a hypothesis that candidates 
found at a mass of 125~GeV are compatible with light CP-even MSSM Higgs
bosons. We show that CEP can give a valuable 
information about spin-parity properties of the Higgs bosons.

\def\thefootnote{\arabic{footnote}}
\setcounter{footnote}{0}

\newpage
